Polarized interfacial tension induces collective migration of cells, as a cluster, in a three-dimensional tissue

S. Okuda[1,*], K. Sato[2,3,**]

[1] Nano Life Science Institute, Kanazawa University, Kanazawa, Japan
[2] Research Institute for Electronic Science, Global Institution for Collaborative Research and Education, Hokkaido University, Sapporo, Japan
[3] Global Station for Soft Matter, Global Institution for Collaborative Research and Education, Hokkaido University, Sapporo, Japan

* satokuda@staff.kanazawa-u.ac.jp
** katsuhiko_sato@es.hokudai.ac.jp

## Abstract (600 characters)
Cells collectively migrate as a cluster in three-dimensional (3D) tissues, such as in embryogenesis and cancer invasion. Here, numerical simulations using a 3D vertex model show that polarized interfacial tension, expressing cell adhesion and cortex contractility, induces the cluster migration in the 3D space. The mechanism is that polarized interfacial tension induced a directional flow of cell-cell interfaces from the front to rear within the whole cluster, producing a driving force, i.e., cells move forward as a cluster by simply expanding and contracting cell-cell boundaries.

## Main text

Cells migrate directionally in three-dimensional (3D) tissues, in which cells are confined from all sides by a dense extracellular matrix (ECM) or tightly adhering cells. These migration characteristics are observed in physiological processes, such as development, immune defense, and wound healing (1-5), as well as in cancer invasion (1-4). Importantly, most cells move forward collectively while forming a cluster in 3D space (which is referred to hereafter as cluster migration). Over the last few decades, there has been intensive study of how cells migrate either individually or collectively on planar substrates (6-8). In particular, recent pioneering studies have revealed a mechanism of single-cell migration in 3D space (9-10); however, little is known about how cells affect cluster migration. Thus, it is unclear under what conditions single-cell or cluster migration is induced.

During cluster migration, cells exert forces on one another. Their collective movement requires interplay of several kinds of force among the cells. Key among these forces may be the active forces commonly involved in a wide range of cell behaviors, i.e., (i) contractile force at the cell-cell boundary by which cells can reduce the boundary area and (ii) adhesion force, which can increase the boundary area and strengthen the cell-cell contact. Although forces (i) and (ii) are expected to be somehow involved in migration (6-8), it is unclear how they are integrated to drive cluster migration in 3D space.

The roles of forces (i) and (ii) have been well investigated in the context of morphogenesis. These studies have used mathematical models, such as Potts models and vertex models (11-12), to explain various phenomena caused by the forces (i)-(ii). Importantly, in those models, forces (i) and (ii) are expressed by interfacial tension in a coarsely-grained manner. For example, actomyosin contractility and cadherin-mediated adhesion are expressed by only one parameter: the strength of interfacial tension.

In addition to forces (i) and (ii), cell polarity is necessary for cell movement to be directional. For example, pioneering studies have shown that each migrating cell is polarized such that actomyosins accumulate locally in the rear part of the cell cortex, biasing cortical tension (9). The polarized tension causes the flow of cell cortex from the front to the rear, generating a propulsive force that moves the cell forward in 3D space (10). Therefore, polarized tension may also play key roles in cluster migration.

In this letter, by developing a simple mathematical model, we clarified whether polarized interfacial tension induces cluster migration. We performed numerical simulations using a 3D vertex model that describes multicellular dynamics at single-cell resolution (13-14). First, we tested whether polarized interfacial tension allows cells to migrate individually in 3D tissue. Second, we clarified that polarized interfacial tension enables cells to migrate collectively as a cluster. Importantly, the mechanism of cluster migration was explained by the directional flow of cell-cell interfaces within the cluster. Moreover, several migratory modes were induced, depending on the strength of adhesion and noise (i.e., cells migrate either as single cells, as a cluster, or aligned like beads on a string), where the migration velocity of cells in a line is higher than that of single cells or of a cluster.

A simple system was considered in which cells are packed within a box [Fig. 1 (a)]. The box is set to a cube within $-L/2 \leq \alpha \leq L/2$ ($\alpha = x, y, z$), where the periodic boundary condition is imposed on $\alpha = \pm L/2$. The box is filled by $N_t$ ($= 432$) cells, in which cells around $y = \pm L/2$ and $z = \pm L/2$ are fixed on the coordinates. Each cell volume is constrained to $V_0$, from which the box size is written as $L = (N_t V_0)^{1/3}$. There are two types of cells in this system, i.e., cells with and without polarity (referred to hereafter as polar and nonpolar cells, respectively). Only polar cells may have polarity and noise to localize actomyosins within the cell, while nonpolar cells have neither polarity nor noise. The nonpolar entities can represent passive, viscoelastic matter such as ECM surrounding the polar cells, in addition to simply representing surrounding cells of a different type than the polar cells.

In the 3D vertex model, the shape of each cell embedded in the 3D tissue is represented by a polyhedron whose vertices are shared by neighboring cells [Fig. 1 (b)]. The shape and configuration of cells are described by the locations of vertices comprising cellular polyhedrons and the topological network among vertices. The time evolution of the $i^{th}$ vertex location, represented by $\boldsymbol{r}_i$, is given by

$$\eta_v \left( \frac{\partial \boldsymbol{r}_i}{\partial t} - \overline{\boldsymbol{v}}_i \right) = -\boldsymbol{\nabla} U. \tag{1}$$

The left side of Eq. (1) is a friction force on the $i^{th}$ vertex (15). Here, $\eta_v$ is the friction coefficient of vertices. Vector $\overline{\boldsymbol{v}}_i$ is a velocity field, defined as the mean velocity of the surrounding cells, where the velocity of the $j^{th}$ cell is defined as the mean velocity of the vertices composing the $j^{th}$ cell. The right side of Eq. (1) is a mechanical force on the $i^{th}$ vertex derived from the effective energy, represented by $U$. During the cell movements described by Eq. (1), individual edges and polygons in the network occasionally shrink to meet or retract from neighboring cells (13-14).

We assume that cells have a simple free energy, given by

$$U \equiv \sum_i^{\text{cell}} \frac{1}{2} k \left( \frac{V_i}{V_0} - 1 \right)^2 + \sum_{\langle ij \rangle}^{\text{boundary}} \gamma_{\langle ij \rangle} A_{\langle ij \rangle}, \tag{2}$$

where the first and second terms indicate volume constraint energy and interfacial energy, respectively. In the first term, the incompressibility of each cell volume is assumed as $V_i \sim V_0$, by which constant $k$ is set much higher than the interfacial energy. In the second term, $\gamma_{\langle ij \rangle}$ and $A_{\langle ij \rangle}$ describe the tension and area of the interface between the $i^{th}$ and $j^{th}$ cells [Fig. 1 (b)]. The resultant of the cortical forces (e.g., actomyosin contractility and cadherin-mediated adhesion) is expressed by tension $\gamma_{\langle ij \rangle}$, e.g., high contractility on the $\langle ij \rangle^{th}$ interface, driven by highly localized actomyosins, is expressed by large $\gamma_{\langle ij \rangle}$. Similarly, adhesion between the $i^{th}$ and $j^{th}$ cells, driven by highly localized adhesion molecules, is expressed by small $\gamma_{\langle ij \rangle}$.

The direction of polarity of each polar cell was simply aligned to the $x$-axis. An interfacial energy that depends on the direction of polarity was introduced as an actomyosin-dependent driving force. The amount of this energy depends on $\theta_{ij}$, the angle from the $x$-axis to the normal vector of each cell surface, i.e., the $\langle ij \rangle^{th}$ boundary, where the direction of the normal vector is toward the outside of the $i^{th}$ cell [Fig. 1 (b)].

In addition, fluctuation of the interfacial energy was introduced only into polar cell surfaces, reflecting the dynamic flow of actomyosin accumulations that is experimentally observed in migrating cells (16-17). Furthermore, adhesion among cells was introduced to the model to represent cell-cell interactions, such as cadherin-mediated adhesions. The amplitude of cell-cell adhesions can differ according to the cell types that interact (18), i.e., polar vs. nonpolar cells. Therefore, we give the interfacial tension on the $\langle ij \rangle^{th}$ boundary in Eq. (2) as

$$\gamma_{\langle ij \rangle} \equiv \frac{\gamma_i^{(p)}}{2}\big(1 + \cos(\theta_{ij} - \pi)\big) + \frac{\gamma_j^{(p)}}{2}\big(1 + \cos(\theta_{ji} - \pi)\big) + \gamma_{\langle ij \rangle}^{(n)} w_{\langle ij \rangle} + \gamma_{\langle ij \rangle}^{(a)}. \qquad (3)$$

The first and second terms indicate polarized interfacial tensions of the $i^{th}$ and $j^{th}$ cells, respectively, expressing the cell polarity. Here, the constants $\gamma_i^{(p)}$ and $\gamma_j^{(p)}$ are coefficients that represent the strengths of polarity of the $i^{th}$ and $j^{th}$ cells, respectively. The polarized tension becomes higher at the rear of each cell, because $\theta_{ij}$ becomes larger toward the rear. Each of the first two terms is maximized ($= \gamma_i^{(p)}$) when $\theta_{ij} = \pi$. The third term denotes noise with a strength $\gamma_{\langle ij \rangle}^{(n)}$, where $w_\alpha$ satisfies $\langle w_\alpha(t) \rangle = 0$ and $\langle w_\alpha(t_1) w_\beta(t_2) \rangle = \delta_\alpha \delta_\beta \exp(-|t_1 - t_2|/\tau)$. $\delta_\alpha$ is the Kronecker delta, and $\tau$ is relaxation time of the noise, as used in previous 2D and 3D models of cell flow (19,20). The fourth term $\gamma_{\langle ij \rangle}^{(a)}$ denotes the strength of adhesion between the $i^{th}$ and $j^{th}$ cells, i.e., lower $\gamma_{\langle ij \rangle}^{(a)}$ indicates higher adhesion strength.

In the interfacial tension of Eq. (3), polarity and noise terms were assigned only to polar cell surfaces, i.e., noise strength was set to $\gamma_{\langle ij \rangle}^{(n)} = \gamma_n$ when either the $i^{th}$ or $j^{th}$ cell is a polar cell, otherwise $\gamma_{\langle ij \rangle}^{(n)} = 0$; and $\gamma_i^{(p)} = \gamma_p$ when the $i^{th}$ cell is a polar cell, otherwise $\gamma_i^{(p)} = 0$. Moreover, adhesions between polar and nonpolar cells can be lower than those between cells of the same type, i.e., the adhesion strength was set to $\gamma_{\langle ij \rangle}^{(a)} = \gamma_0$ between cells of the same type, and to $\gamma_{\langle ij \rangle}^{(a)} = \gamma_1$ ($\geq \gamma_0$) between polar and nonpolar cells. Parameters are nondimensionalized by unit length $(V_0)^{1/3}$, unit time $\eta_c/20\gamma_0(V_0)^{2/3}$, and unit energy $\gamma_0(V)^{2/3}$. Relaxation time was set to $\tau = 3\eta_c/20\gamma_0(V_0)^{2/3}$. Remaining parameters are $\gamma_1$, $\gamma_p$, and $\gamma_n$.

First, we tested whether cells move as individual cells, and found that the interfacial tension of Eq. (3) succeeded in inducing single-cell migration for appropriate parameter values (Fig. 2). Here, cells migrated in cases with or without noise [Fig. 2 (a)], where migrations in the case with noise required lower polarity than those without noise. During single-cell migration, a cell moved directionally forward while sequentially rearranging its contiguity [Fig. 2 (b), Video 1]. In this process, the surface of each cell flowed from the front to the rear of the cell, and this directional flow was caused by the polarized surface tension [Fig. 2 (c), Video 2], e.g., an edge connected to the cell near its front end was reconnected to a new cell face, C. Face C moved from the front to the rear of the cell as the interfacial tension was increased. Simultaneously, face C expanded in the first half ($t$ = 60 to 160) and shrank in the second half ($t$ = 240 to 360). Face C eventually reached the rear of the cell, and disappeared by reconnecting to a new edge. The directional flow of cell-cell interfaces repeated cyclically, driving individual cells forward.

The mechanism of single-cell migration observed in this study is essentially the same as that proposed in previous works (8-10), in which polarized interfacial tension causes the surface to flow within each cell from front to rear, and is the driving force of single-cell movement. A difference is that our model is more abstract but expresses the effects of both cortical tension and cell-cell adhesion using the polarized interfacial tension.

Next, we increased the number of polar cells in the box to assess whether clusters of cells move collectively (Fig. 3). When $\gamma_{\langle ij \rangle}^{(a)}$ had the same amplitude among all cells, polar cells moved forward as individual cells. On the other hand, when $\gamma_{\langle ij \rangle}^{(a)}$ was greater between polar and nonpolar cells, i.e., when cells of the same type tended to aggregate, polar cells formed a cluster that was maintained during unidirectional movement [Fig. 3 (a), Video 3].

Interestingly, the mechanism of cluster migration was analogous to that of single-cell migration. That is, single-cell migration was induced by the flow of interfaces within each cell, while cluster migration was induced by the flow of interfaces within the whole cluster. In other words, the interfaces between polar and nonpolar cells relayed along the cluster surface from cells near the front to cells near the rear of the cluster [Fig. 3 (b)].

To describe the process more explicitly, we focus on face S, located at a boundary between polar and nonpolar cells [Fig. 3 (b)]. Face S was initially located on cell (i) of the cluster. In the simulation, face S spread to the adjacent cell (ii) located behind the cell (i), gradually moved from cell (i) to cell (ii), and eventually moved to cell (ii) completely.

The relay of interfaces from cell to cell can be understood by considering the difference in interfacial tensions between neighboring cells. Let us focus on the situation in Fig. 3 (b) at $t = 440$, in which face S lies across both cells (i) and (ii). Face S on cell (i) had a large interfacial tension because it was located at the rear of the cell (i) and was directed toward the negative $x$-axis. In contrast, the face S portion of cell (ii) had a small interfacial tension because it was located at the front of the cell (ii) and was directed toward the positive $x$-axis. Because interfacial tension is equivalent to energy density, face S energetically preferred to be located on cell (ii) rather than cell (i). Although similar movements occur at the boundaries between polar cells, the movements between neighboring polar cells compete to fix the configuration of polar cells inside the cluster. The cell-to-cell relay of interfaces continued from the front-end to the rear-end of the cluster, and this directional flow was cyclically repeated to move the cluster forward. These results illustrate a common mechanism for both single-cell and clustered cell migrations, i.e., by regarding either a single cell or a cell cluster as an object, the object moves forward by the directional flow of the object surface from the front to the rear [Fig. 3 (c)]. By providing a simple mathematical model, we analytically showed that polarized interfacial tension certainly induces single-cell and cluster migrations (Appendix).

Additionally, to analyze what conditions determine single-cell and cluster migrations, we calculated the effects of each term in Eq. (3) on cell behaviors, and found four characteristic states [Fig. 4 (a), Video 3-5], as follows. I). In the case with low polarity ($\gamma_n/\gamma_0 < 0.25 \sim 0.5$), cells were nearly arrested in the coordinates, and did not migrate. When the polarity ($\gamma_p$) was increased, the cell state changed from nonmigrating to migrating, and then the migration velocity increased. II) In the case with high noise ($\gamma_n/\gamma_0 = 1.25$) and homogenous adhesion among cells ($\gamma_1 = \gamma_0$), cells migrated as individual cells. III) When the adhesion between polar and nonpolar cells was lower than between the same type of cells ($\gamma_1 > \gamma_0$), polar cells formed an aggregate. IV) Interestingly, in the case without heterogenous adhesion ($\gamma_1 = \gamma_0$), cells aligned like a string of beads. More quantitatively, the cell state changed from nonmigrating to migrating in a nonlinear manner, and the nonlinearity was reduced by an increase in the amplitude of noise [Fig. 4 (b)]. Moreover, the cluster size increased as the adhesion between polar and nonpolar cells decreased ($\gamma_1$ increased), whereas the migration velocity decreased [Fig. 4 (c)]. Notably, the mean velocity of migrating cells was accelerated simply by increasing the number of polar cells from one cell to multiple cells [Fig. 4 (d)].

In summary, simulations using the 3D vertex model showed that polarized interfacial tension induces collective cell migration as a cluster in 3D space. We explained the mechanism of cluster migration by analogy to the Marangoni effect, i.e., interfacial tension induced the directional flow of cell-cell interfaces within the cell cluster. We also identified the conditions that determine single-cell and cluster migrations, i.e., the existence of cell polarity to directionally move cells forward, and heterogenous adhesion among cells to form a cluster. Because cell-cell interfacial tension can be regulated by the accumulation of actomyosin and adhesion molecules, both single-cell and cluster migrations can be thus driven by their localized accumulations within each cell. Whether migrating cells form a cluster may be determined by the type and expression level of adhesion molecules among the cells.

Although our results served to clarify the mechanism of cluster migration, several further physical questions remain, e.g., how migrating cells aligned themselves like beads on a string despite the absence of heterogenous adhesion among cells [Fig. 4 (a)], and how the migration velocities were accelerated by increasing the number of cells [Fig. 4 (d)]. More detailed analyses are needed to address those issues and others, e.g., the geometry of packed cells imposes energy barriers to cell rearrangements (21) that must be overcome for cell clusters to migrate in 3D space. From our results, it is evident that cells did sequentially cross barriers in the 3D packing geometry; however, the process of crossing the energy barriers is still unclear. Spatial dimensions may be important for cells to cross these energy barriers, such as topological dimensions and Poisson effects (22, 23). These could be clarified by detailed comparisons between 2D and 3D analyses.

From a biological point of view, the migration modes and state transitions found in the simulation may be worth discussing, i.e., cells migrate as single cells, as a cluster, or aligned in a row [Fig. 4 (a)]. Similar modes are observed in cancer metastasis, where cells migrate as single cells or forming clusters, alveolar, or trabecular structures (the latter corresponding to the aligned migration mode) (24). In our results, the migration velocity of the line of cells was higher than that of the cluster [Fig. 4 (a)], which corresponds to the invasive potential of cancer cells (24). These results raise a biological question, i.e., whether collective cell migration in living systems is driven by interfacial tension.

## Acknowledgements
We thank Tetsuya Hiraiwa at Mechanobiology Institute, National University of Singapore, and Masanobu Oshima at Cancer Research Institute, Kanazawa University for discussions.

This work was supported by the Japan Science and Technology Agency (JST), CREST Grant No. JPMJCR1921 (S.O.), PRESTO Grant No. JPMJPR16F3 (S.O.); the Japan Society for the Promotion of Science (JSPS), KAKENHI Grants No. 21H01209 (S.O.), 20K20958 (S.O.), 19H04777 (S.O.), 20K03871 (K.S.), 18H01135 (K.S.), and 17KT0021 (S.O.); Global Station for Soft Matter at Hokkaido University (K.S.) ; the Research Program of "Five-star Alliance" in "NJRC Mater. & Dev." (K.S); the Uehara Memorial Foundation, Japan (S.O.); the NOVARTIS Foundation (Japan) for the Promotion of Science (S.O.); the Brain Science Foundation (S.O.); and the World Premier International Research Center Initiative, Ministry of Education, Culture, Sports, Science and Technology (MEXT), Japan (S.O.).

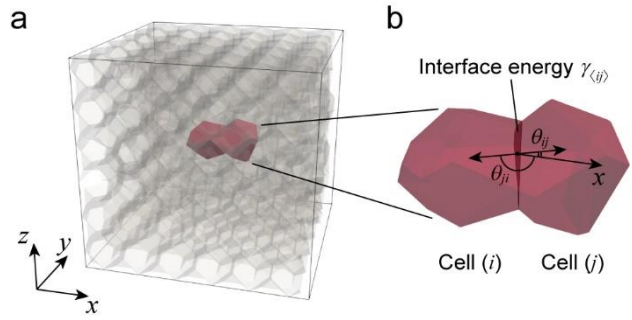

FIG. 1 Setup of the model. **(a)**. System box filled by cells in 3D space. Individual cell shapes are represented by polyhedrons. **(b)**. Cell-cell interfacial energy, $\gamma_{\langle ij \rangle}$, is introduced into the $\langle ij \rangle$th polygonal boundary face shared by the $i$th and $j$th adjacent cells. The angle for the $i$th cell to the $\langle ij \rangle$th face, $\theta_{ij}$, is defined as that from the x-axis to the normal vector of the $\langle ij \rangle$th face, whose direction is toward the outside of the $i$th cell.

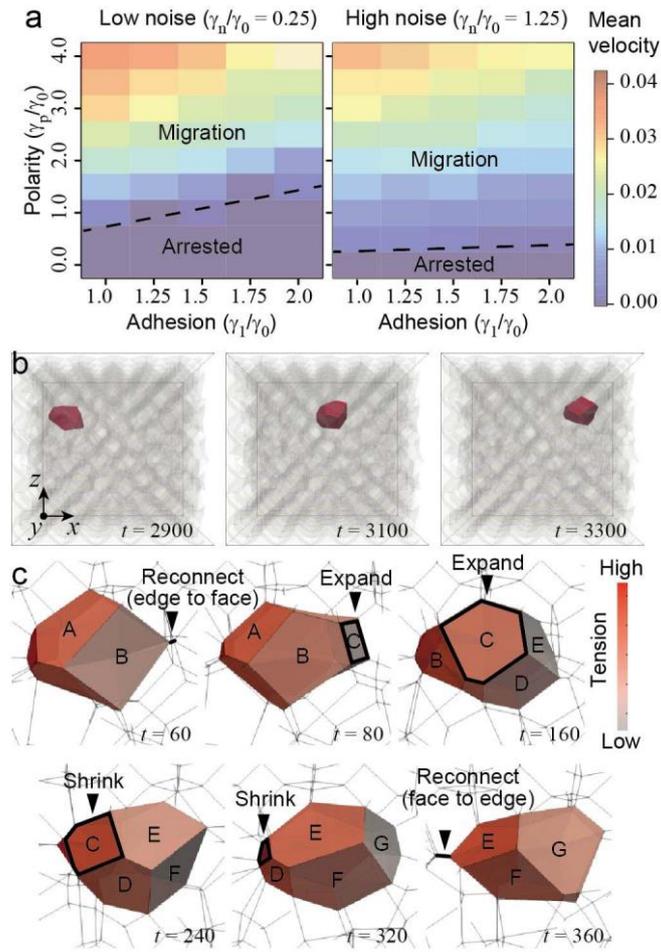

FIG. 2. Single-cell migration in 3D space. (a). State and velocity diagrams as a function of polarity ($\gamma_p$), adhesion ($\gamma_1$), and noise ($\gamma_n$). (b). Development of single-cell migration over time (also shown in Video 1). A single polar cell is colored in red, and nonpolar cells are colored in transparent gray. (c). Geometry and tension of cell-cell interfaces during single-cell migration (also shown in Video 2). The color scale at right indicates interfacial tensions of individual surfaces of the single polar cell, some of which are tagged alphabetically (A through G).

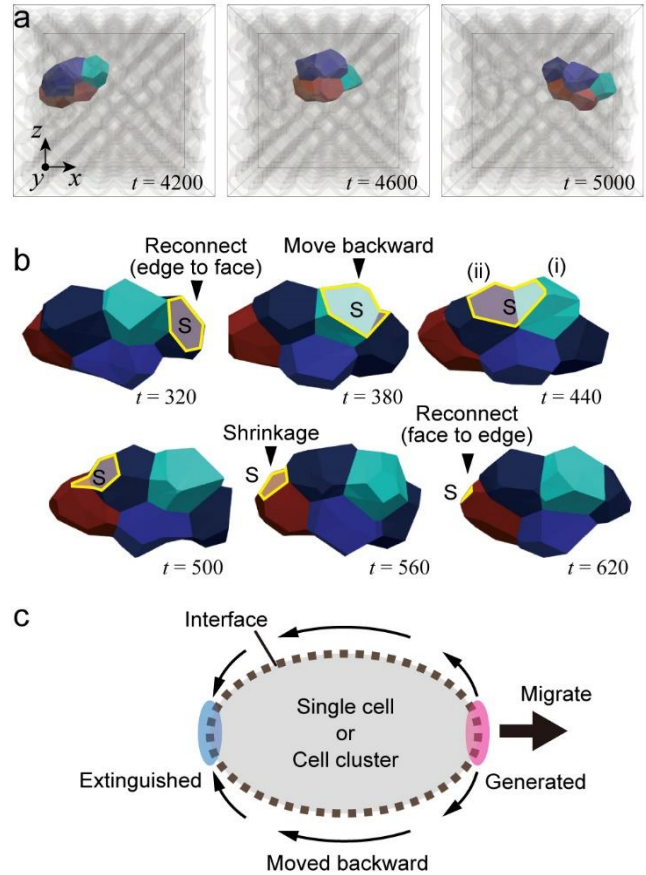

FIG. 3. Cluster migration in 3D space. **(a)**. Development of clustered cell migration over time (also shown in Video 3). Individual polar cells are designated with different colors, and nonpolar cells are colored in translucent gray. **(b)**. Relay process of a boundary face from the front to the rear within a cluster. The face S at the boundary between polar and nonpolar cells was relayed from the front end to the rear of the cluster. At $t = 440$, face S lies across cells (i) and (ii). **(c)**. Schematic illustration of the common mechanism of single-cell and cluster migrations. In both cases, new interfaces are sequentially generated at the front, moved backward, and extinguished in the rear. The unidirectional flow of cell-cell interface generates a driving force.

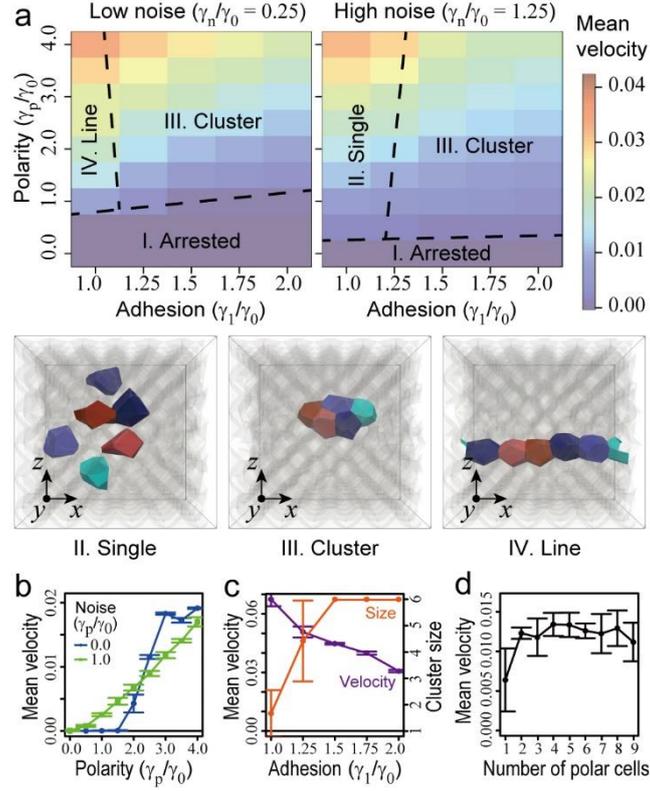

FIG. 4. Dependence of 3D cell migration on components of interfacial tension. (a). State and velocity diagrams as a function of polarity ($\gamma_p$), adhesion ($\gamma_1$), and noise ($\gamma_n$). Snapshots of migrations as single cells (II), a cluster (III), and cells in alignment (IV) are shown below. The development of these migration modes over time are shown in Videos 4, 3, and 5. Individual polar cells are designated with different colors, and nonpolar cells are colored with translucent gray. (b). Mean velocity as a function of polarity and noise. (c). Mean velocity and cluster size as functions of adhesion. (d). Mean velocity as a function of the number of polar cells.